\documentclass[12pt,thmsa]{article}
\usepackage{sw20lart}

\input{tcilatex}
\begin{document}

\begin{center}
{\LARGE An Investigation of Laboratory-Grown ``Ice Spikes'' \bigskip }

\textsc{K. G. Libbrecht and K. Lui}\smallskip \footnote{%
Address correspondence to $kgl@caltech.edu;$ URL: http://www.its.caltech.edu/%
\symbol{126}atomic/}\textsc{\ \smallskip }

\textit{Norman Bridge Laboratory of Physics, California Institute of
Technology 264-33, }

\textit{Pasadena, CA 91125\bigskip }

{\small [Posted October 11, 2003]}
\end{center}

\noindent \textbf{Abstract. We have investigated the formation of 10-50 mm
long ``ice spikes'' that sometimes appear on the free surface of water when
it solidifies. By freezing water under different conditions, we measured the
probability of ice spike formation as a function of: 1) the air temperature
in the freezing chamber, 2) air motion in the freezing chamber (which
promotes evaporative cooling), 3) the quantity of dissolved salts in the
water, and 4) the size, shape, and composing material of the freezing
vessel. We found that the probability of ice spike formation is greatest
when the air temperature is near -7 C, the water is pure, and the air in the
freezing chamber is moving. Even small quantities of dissolved solids
greatly reduce the probability of ice spike formation. Under optimal
conditions, approximately half the ice cubes in an ordinary ice cube tray
will form ice spikes.}

\textbf{Guided by these observations, we have examined the Bally-Dorsey
model for the formation of ice spikes. In this model, the density change
during solidification forces supercooled water up through a hollow ice tube,
where it freezes around the rim to lengthen the tube. We propose that any
dissolved solids in the water will tend to concentrate at the tip of a
growing ice spike and inhibit its growth. This can qualitatively explain the
observation that ice spikes do not readily form using water containing even
small quantities of dissolved solids.}

\section{Introduction}

When water freezes into ice, the expansion that occurs during the
solidification process sometimes causes the formation of ``ice spikes'' that
rise out of the free ice surface. The phenomenon has been observed
sporadically out-of-doors in cold climates for many decades \cite{dorsey,
hallet, abrusci, knight, perry, perry2}. These rare sightings have
documented the appearance of rather large ice spikes, perhaps 100 mm in
length, sometimes with triangular cross-sections, usually forming overnight
in containers of standing water. Much smaller ice spikes (also known as ice
spicules) have been observed coming out of sleet particles \cite{alty,
bally, blanchard, mason}. These appear in droplets of order 1 mm in size or
larger, and they have been successfully reproduced and studied in the
laboratory \cite{blanchard, mason}.

More recently, numerous reports have appeared on the internet describing the
appearance of 10-50 mm long ice spikes forming when water is frozen in
ordinary ice cube trays in ordinary household freezers \cite{webstuff}. To
our knowledge, the appearance of ice spikes under such circumstances has
received little attention in the scientific literature (however see \cite
{perry, perry2}), and the topic has not been the subject of any systematic
laboratory studies. A key element in the production of ice spikes -- the
purity of the water that is frozen -- has only briefly been touched upon in
any of the previous studies of ice spikes \cite{mason}. Ice spikes rarely
form when tap water is frozen, but do so frequently using distilled water.
This important point was discovered and communicated to us by J. F. Cooper
III \cite{cooper}, although it appears it was independently discovered by a
number of others \cite{webstuff}.

If household distilled water is used (the type sold in grocery stores), then
a plastic ice cube tray in a typical freezer will yield approximately 1-4
ice spikes that look like the examples shown in Figure 1. The spikes form at
apparently random angles with respect to vertical and with various lengths.
Occasionally more than one spike will grow out of a single cube, and once we
found three spikes emanating from a common base. The longest ice spike we
observed in any of our experiments using ice cube trays was 56 mm from base
to tip. These spikes typically have roughly circular cross-sections
(sometimes with a rounded triangular cross-section near the base), and the
spikes often display narrow, fairly sharp tips. Even when an ice cube does
not contain a spike, the ice surface is typically not flat, but includes
several mm-scale surface ice lumps. Tap water usually does not form ice
spikes, although it does form ice lumps.

The generally accepted mechanism for the formation of ice spikes of all
sizes is the Bally-Dorsey model \cite{blanchard, bally, dorsey}. In the case
of an ice spike forming in an ice cube tray, water first freezes at the
surface, starting at the edges the cube, and the ice subsequently expands
laterally until only a small hole in the ice surface remains. Then the
continued freezing of water beneath the surface forces water up through the
hole, where it freezes around the edge of the hole to form the beginnings of
a hollow tube. Continued freezing forces water up through the tube, where it
freezes around the rim and lengthens the tube. At some point the tube
freezes shut and growth stops.

Below we examine ice spike formation under a variety of different growth
conditions. Guided by these observations, we have further examined the
Bally-Dorsey model, in particular regarding the growth of ice spikes in the
presence of dissolved solids in the water.

\FRAME{fhFU}{6.0105in}{2.8712in}{0pt}{\Qcb{A number of ice spikes grown in
an ordinary plastic ice cube tray using distilled water. Note the spikes
grow at many different positions and angles, and to various heights up to $%
\sim 50$ mm.}}{}{multspikes2.gif}{\special{language "Scientific Word";type
"GRAPHIC";maintain-aspect-ratio TRUE;display "USEDEF";valid_file "F";width
6.0105in;height 2.8712in;depth 0pt;original-width 899.625pt;original-height
428.375pt;cropleft "0";croptop "1";cropright "1";cropbottom "0";filename
'C:/aaa/Papers/IceSpikes/multspikes2.gif';file-properties "XNPEU";}}

\section{Observations}

Our observations were primarily focused on measuring the probability of ice
spike formation as a function of various growth conditions. The experiments
were performed in an insulated cylindrical copper cold tank measuring 0.5
meters in diameter by 0.6 meters in height that was cooled by circulating
cold methanol through copper tubing soldered to the sides and top of the
tank. Freezing vessels, usually ordinary plastic ice cube trays in which the
volume of a single cube was 18 ml, were placed on a steel wire rack that
held the trays in the middle of the tank. For some trials a fan inside the
tank circulated the air to promote evaporative cooling. The fan was either
on or off and circulated the air in the tank roughly once every 10-20
seconds. We used laboratory deionized water, without further purification,
for all our trials.

Ice spikes form quickly once the water begins to freeze. We videotaped
several trials using a miniature TV camera placed inside the tank and found
that spikes typically grow to their full height in 3-10 minutes, at growth
velocities of roughly $50$ $\mu $m/sec. If sufficiently disturbed, some
spikes were observed to change direction in the middle of their growth. The
growth of a spike stops when the tube freezes shut, which happens long
before the entire ice cube is frozen.

\textbf{Dependence on Temperature and Air Motion.} We froze several thousand
ice cubes (by which we mean those in ordinary ice cube trays, which are not
actually cubical in shape) in several dozen trials to determine the
probability of ice spike formation as a function of temperature and air
motion. By probability we mean the probability that a single ice cube in an
ice-cube tray will produce a spike that rises more than a few millimeters
above the ice surface. We assume that all the cubes in a tray are
independent of one another.

Figure 2(a) shows the probability as a function of temperature for still
air. The error bars in temperature arise because the room-temperature water
initially produced a substantial perturbation of the air temperature in the
tank. The individual temperature error bars on this plot span from the
initial chamber temperature (equal to the lowest temperature during the
trial, also equal to the final chamber temperature) to the temperature a few
minutes after the trays were placed in the chamber (the highest temperature
during the trial). The displayed temperature was the average of these two
values. The vertical error bars were derived assuming Bernoulli statistics.
Figure 2(b) displays similar data, except that for these points the fan
circulated air in the chamber.

The sharp decline in ice spike formation above -5 C coincided with a sudden
increase in freezing time that occurred at about the same temperature.
Substantially below -5 C, it took about two hours for the ice cubes to
freeze, and the freezing time slowly increased with increasing temperature.
Around -5 C the freezing time increased abruptly, and at temperatures above
-4 C the water often reached a metastable supercooled state that did not
freeze for many hours.

The broad maxima seen in Figure 2 suggest that most household freezers are
sufficient to produce ice spikes using distilled water, which we confirmed
with additional trials using a number of freezers around the Caltech campus.
Modern frost-free freezers circulate cold dry air through the freezing
compartment using a fan, in contrast to older freezers that generally do not
have circulating fans. The former are more likely to produce ice spikes, as
we can see from the data in Figure 2.

\FRAME{ftbFU}{6.1255in}{2.418in}{0pt}{\Qcb{Measurments of the probability
that a single ice cube in a tray will produce an ice spike longer than a few
millimeters. Each point refers to a single trial consisting of $\sim 50$ ice
cubes. The left plot is for ice cubes grown in still air and the right plot
is with a fan circulating air in the cold tank. The curves in both plots
were drawn to guide the eye.}}{}{tempdata2.gif}{\special{language
"Scientific Word";type "GRAPHIC";maintain-aspect-ratio TRUE;display
"USEDEF";valid_file "F";width 6.1255in;height 2.418in;depth
0pt;original-width 1066.75pt;original-height 419.3125pt;cropleft "0";croptop
"1";cropright "1";cropbottom "0";filename
'C:/aaa/Papers/IceSpikes/tempdata2.gif';file-properties "XNPEU";}}

\textbf{Dependence on Salt Concentration.} We examined how the probability
of ice spike formation varied with dissolved solids by forming solutions of
ordinary table salt in deionized water at different concentrations and
freezing them at a chamber temperature of -7 C. Figure 3 shows the ice spike
probability as a function of solution concentration. Since medium-hard tap
water is typically a $\sim $10$^{-3}$ molar solution of various mineral
salts, we see that the data in Figure 3 are consistent with the observation
that ice spikes only rarely form from tap water. Even a concentration of 10$%
^{-5}$ molar is sufficient to reduce the ice spike probability by a factor
of two from the value for deionized water.

\FRAME{ftbhFU}{3.8372in}{2.9801in}{0pt}{\Qcb{Probability of ice spike
formation as a function of the solution concentration of table salt in
dionized water.}}{}{salt.gif}{\special{language "Scientific Word";type
"GRAPHIC";maintain-aspect-ratio TRUE;display "USEDEF";valid_file "F";width
3.8372in;height 2.9801in;depth 0pt;original-width 544.3125pt;original-height
422.3125pt;cropleft "0";croptop "1";cropright "1";cropbottom "0";filename
'C:/aaa/Papers/IceSpikes/salt.gif';file-properties "XNPEU";}}

\textbf{Dependence on Container Size and Composition.} We performed several
trials near the temperature peak (see Figure 2) using old-fashioned aluminum
ice cube trays, after the individual cubes had been sealed using silicone
caulk (to prevent the flow of water between cubes). These yielded few
spikes, suggesting that the thermally insulating properties of plastic ice
cube trays are beneficial. We therefore did some additional trials using
plastic ice cube trays to which we had added additional styrofoam insulation
on the bottom. These trials indicated that the added insulation reduced the
ice spike probability by a small but significant factor when the fan was on.
These observations indicate that an ordinary plastic ice cube tray provides
a near optimal amount of thermal insulation for forming ice spikes.

We also performed several trials using cylindrical acrylic freezing vessels
of different sizes, ranging from approximately 1 cm$^{3}$ to 25 cm$^{3}$ in
volume. In these trials we observed that the larger containers had a
slightly higher probability of ice spike formation, and that the length and
girth of the ice spikes is somewhat larger in the larger containers.
Ultra-pure water increases the production of ice spikes in larger
containers, allowing the formation of spikes up to 100 mm in length \cite
{elgersma}.

\FRAME{fthFU}{4.6596in}{3.7403in}{0pt}{\Qcb{The Bally-Dorsey model for ice
spike formation. Ice spikes grow as the solidification process forces water
up through an ice tube, where it freezes and lengthens the tube. As
described in the text, this model can explain why ice spikes form
preferentially in pure water and in the presence of air flow.}}{}{%
icespikeformation.jpg}{\special{language "Scientific Word";type
"GRAPHIC";maintain-aspect-ratio TRUE;display "USEDEF";valid_file "F";width
4.6596in;height 3.7403in;depth 0pt;original-width 417.8125pt;original-height
335pt;cropleft "0";croptop "1";cropright "1";cropbottom "0";filename
'C:/aaa/Papers/IceSpikes/icespikeformation.jpg';file-properties "XNPEU";}}

\section{Discussion}

The Bally-Dorsey model for ice spike formation is illustrated in Figure 4.
If the chamber temperature is sufficiently cold (say around -10 C), then
nucleation sites are plentiful and ice first forms on the surface near the
edges of the container. Ice subsequently grows in from the edges until only
a small hole in the ice sheet remains unfrozen. If the freezing rate is
sufficiently high beneath the surface, then the expansion that accompanies
solidification will begin to force water out through the remaining hole in
the ice. One frequently observes partially frozen ice cubes in which the ice
surface has been wet by this process. If the initial freezing is slow enough
that the growth is primarily limited by attachment kinetics, then the ice
will form faceted structures that can result in the formation of triangular
holes and ice spikes with triangular cross-sections.

If conditions are right, the water pushed through the hole in the ice will
freeze around the edges of the hole and form a short tube of ice. Once a
tube is established, water forced up through the tube will form an unfrozen
droplet perched atop the tube. The droplet will freeze around the edges of
the tube, thus increasing the length of the tube.

We can see that the growth of an ice spike will be stable to small
perturbations in the flow of water up the tube. If the water flow increases,
then the unfrozen droplet on top of the tube will increase in size. The
larger droplet has a larger surface area, so evaporative cooling will tend
to increase the rate of freezing around the edges of the tube. With a large
enough flow, water would spill over the sides of the tube, but ice spikes
typically show no signs of this. For a reduced rate of water flowing up the
tube, the droplet will shrink and the freezing rate will diminish. This
suggests that, over some range of water flows, the freezing at the top of
the tube will adjust to match the water flow from below. The tube will
continue growing until it freezes shut somewhere along its length.

The dramatic dependence of ice spike formation on impurities in the water
likely stems from the fact that impurities are generally not incorporated
into the ice lattice during the freezing process. Thus, during the formation
of an ice spike, impurities will become increasingly concentrated in the
small unfrozen droplet at the top of the tube. These impurities reduce the
freezing rate and so the growth of the tube. When the initial impurity rate
is sufficiently high, most nascent tubes grow slowly enough that they freeze
shut before reaching an appreciable length.

In the rare occasions when exceptionally large spikes grow in natural,
outdoor ice formations, it appears some other mechanism may be necessary to
remove the impurities that build up at the top of the growing tube. The
impurities may be forced into pockets that freeze more slowly, as happens
with sea ice, or perhaps a convective flow replaces the water at the top of
the tube with fresh water from below. Significant convection is unlikely in
the smaller tubes that form in ice cube trays.

This model also provides a simple explanation for the fact that air motion
enhances the production of ice spikes. Evaporative cooling will enhance the
stabilization mechanism described above, and it will increase the freezing
rate of the droplet on top of a growing tube. Since evaporative cooling will
not greatly increase the freezing rate of the inner walls of the tube, an
increase in the freezing rate at the top should produce longer ice spikes.

It appears that evaporative cooling plays an important role in ice spike
formation even in still air. If evaporative cooling were absent, then
conductive heat flow considerations suggest that the ice growth rate inside
the tube would not be dramatically slower than that at the top of the tube.
If that were the case, then the tubes would freeze shut relatively quickly
and one would not expect ice spikes that are 10-20 times longer than their
diameters, as we observe.

It may also be the case that the water flow through the tubes slows the
freezing rate along the tube walls by a substantially amount, inasmuch as it
affects the heat flow inside the tube. We might expect such a mechanism to
work less well at lower temperatures, and this might explain the decrease in
ice spike formation below the peaks seen in Figure 2.

Making the Bally-Dorsey model more quantitative remains challenging. One
would need to model heat flow in and around the tube, evaporative cooling
from the spike tip, the degree of supercooling at various points along the
tube, the effect of flow on freezing, and many other factors. It should be
possible, however, to manufacture ice spikes in a much more controlled
manner than has been done to date. Once a method is developed to begin the
ice tube, water could be forced up through the growing tube using an
externally applied pressure. One could then measure the tube growth rates
under different conditions to explore this fascinating phenomenon in more
detail.

\end{document}